\documentclass[a4paper,twocolumn]{article}
\usepackage{arxiv}
\usepackage{cite}
\usepackage{amsmath,amssymb,amsfonts}
\usepackage{algorithmic}
\usepackage{graphicx}
\usepackage{textcomp}

\title{A Neural Network Model of the Entorhinal Cortex and Hippocampus for Event-order Memory Processing}

\author{Hiroki Nakagawa$^{1}$, Katsumi Tateno$^{1, 2}$, Kensuke Takada$^{1}$, Takashi Morie$^{1, 2}$\\
	${}^{1}$ Graduate School of Life Science and Systems Engineering, Kyushu Institute of Technology\\
	2-4, Hibikino, Wakamatsu-ku, Kitakyushu, 808-0196, Japan\\
	${}^{2}$  Research Center for Neuromorphic AI Hardware\\
	2-4, Hibikino, Wakamatsu-ku, Kitakyushu, 808-0196, Japan\\
	Corresponding author: Katsumi Tateno~(e-mail: \texttt{tateno@brain.kyutech.ac.jp})\\
}

\begin{document}
\twocolumn[
This work has been submitted to the IEEE for possible publication. Copyright may be transferred without notice, after which this version may no longer be accessible.
\maketitle
\begin{abstract}
To solve a navigation task based on experiences, we need a mechanism to associate places with objects and to recall them along the course of action. In a reward-oriented task, if the route to a reward location is simulated in mind after experiencing it once, it might be possible that the reward is gained efficiently. One way to solve it is to incorporate a biologically plausible mechanism. In this study, we propose a neural network that stores a sequence of events associated with a reward. The proposed network recalls the reward location by tracing them in its mind in order. We simulated a virtual mouse that explores a figure-eight maze and recalls the route to the reward location. During the learning period, a sequence of events relating to firing along a passage was temporarily stored in the heteroassociative network, and the sequence of events is consolidated in the synaptic weight matrix when a reward is fed. For retrieval, the sequential activation of conjunctive cue-place cells toward the reward location is internally generated by an impetus input. In the figure-eight maze task, the location of the reward was estimated by mind travel, irrespective of whether the reward is in the counterclockwise or distant clockwise route. The mechanism of efficiently reaching the goal by mind travel in the brain based on experiences is beneficial for mobile service robots that perform autonomous navigation.
\end{abstract}
\vspace{1em}
]

\section{Introduction}
Robot technology has made remarkable progress, but it has yet to reach the point where robots can interact with humans on an equal footing. Home service robots are expected to learn about an event and accumulate experiences as an episode. An environment is considered different from house to house and from person to person. A circumstance may be less likely to be repeated. Currently, artificial intelligence is mainly aimed at generalizing and processing a large amount of information and an individual preference is unconsidered. Several studies relating to neural networks for robots that perform services in human living spaces have emerged~\cite{RIZZI201732, Cazin2019, Tanaka2020}.

Humans tend to memorize impressive events they encountered in the past, and they tend to retrieve a series of experienced events in a certain circumstance. Storage of information expressing when, where, and what we did is called episodic memory. Each series of episodes is considered to be linked intermittently~\cite{tulving1972episodic}. The medial temporal lobe contributes to episodic memory. Any damage to the medial temporal lobe causes severe impairment of episodic memory formation~\cite{scoville1957loss}.

Episodic memory-related neurons, such as place cells~\cite{o1978hippocampus} and time cells~\cite{macdonald2011hippocampal}, have been found in the medial temporal lobe of mice and rats. Place cells in the hippocampus fire at a specific location~\cite{o1971hippocampus}. Tolman~\cite{tolman1948cognitive} introduced a cognitive map: the idea that mice or rats form maps in their brain as they move through space. In discovering place cells, O’Keefe et al.~\cite{o1978hippocampus} proposed that a cognitive map exists in the hippocampus. Place cells are considered a component of the cognitive map. Speed cells~\cite{kropff2015speed} and head-direction cells~\cite{taube1990head} are also found in the medial temporal lobe. Speed cells in the medial entorhinal cortex represent movement speed. The self-position can be estimated by performing path integration based on the input from these cells~\cite{gothard1996dynamics}.

Synaptic plasticity, a function that enhances or attenuates synapses by cell activity, is deeply involved in episode memory. Information transmission in the brain is performed via synapses, and various learning rules have been proposed based on this idea of synaptic plasticity. Hebb’s rule is a typical learning rule of synaptic plasticity~\cite{hebb2005organization}. Synaptic plasticity that depends on the relative timing between presynaptic and postsynaptic neurons is called spike-timing-dependent plasticity (STDP). However, the time window for the Hebb’s rule is narrow. The interval between presynaptic and postsynaptic spikes must be within a few tens of milliseconds. Memorizing episodes involves remembering events that happened at the same time and those that happened in sequence over time. Events that occur more than a few tens of seconds apart are also linked by the hippocampus as episodic memories. For learning, there is also a distant reward problem. Typically, there is a couple of minutes of the temporal gap between the neural activity that determined the action and reward. To solve the distal reward problem, we need a mechanism that ties events together.

In this study, we propose a neural network model that consolidates a temporal order of cue-place association. A physiology-based simplified model assuming the entorhinal cortex and hippocampus is used to associate cues with places. Additionally, as a part other than the hippocampus, we propose a long-term memory matrix in which a sequential relationship of event occurrence is preserved over time. Furthermore, we use computer simulations to verify whether the proposed model retrieves the order of the association information of places with objects. 

\section{Neural network model}

\begin{figure}[h]
  \begin{center}
    \includegraphics[width=\linewidth, keepaspectratio, clip]{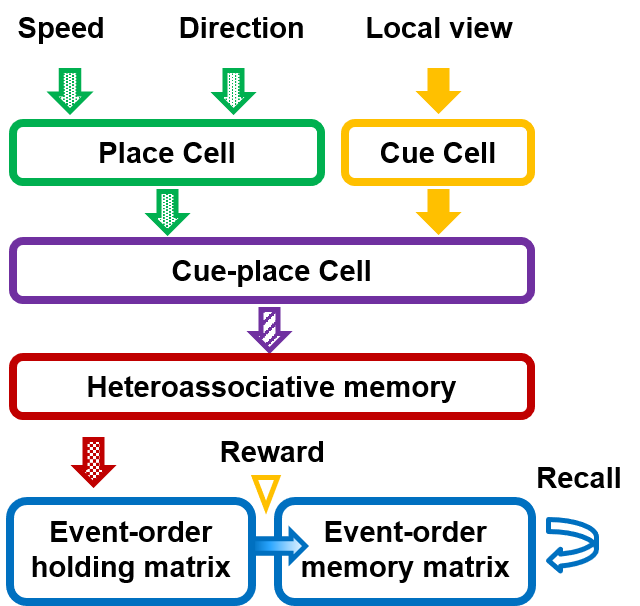}
  \end{center}
    \caption{Proposed network model.}
    \label{fig:input-from-outside}
\end{figure}

Fig.~\ref{fig:input-from-outside} depicts a connection diagram of the proposed neural network model. Signals of own speed and head orientation and an external cue signal are inputted into the neural network. The speed signal and head-direction signal are integrated into place cells. The cue signal triggers cue cells. The outputs of the place and cue cells are integrated into cue-place cells. In this study, the integrated cue-place information was treated as an event. The events are stored along with the movement of the passage.  Neuronal activities of proximal cue-place cells are associated with a sequence of events in the heteroassociative network.

\begin{figure}[h]
  \begin{center}
    \includegraphics[width=\linewidth, keepaspectratio, clip]{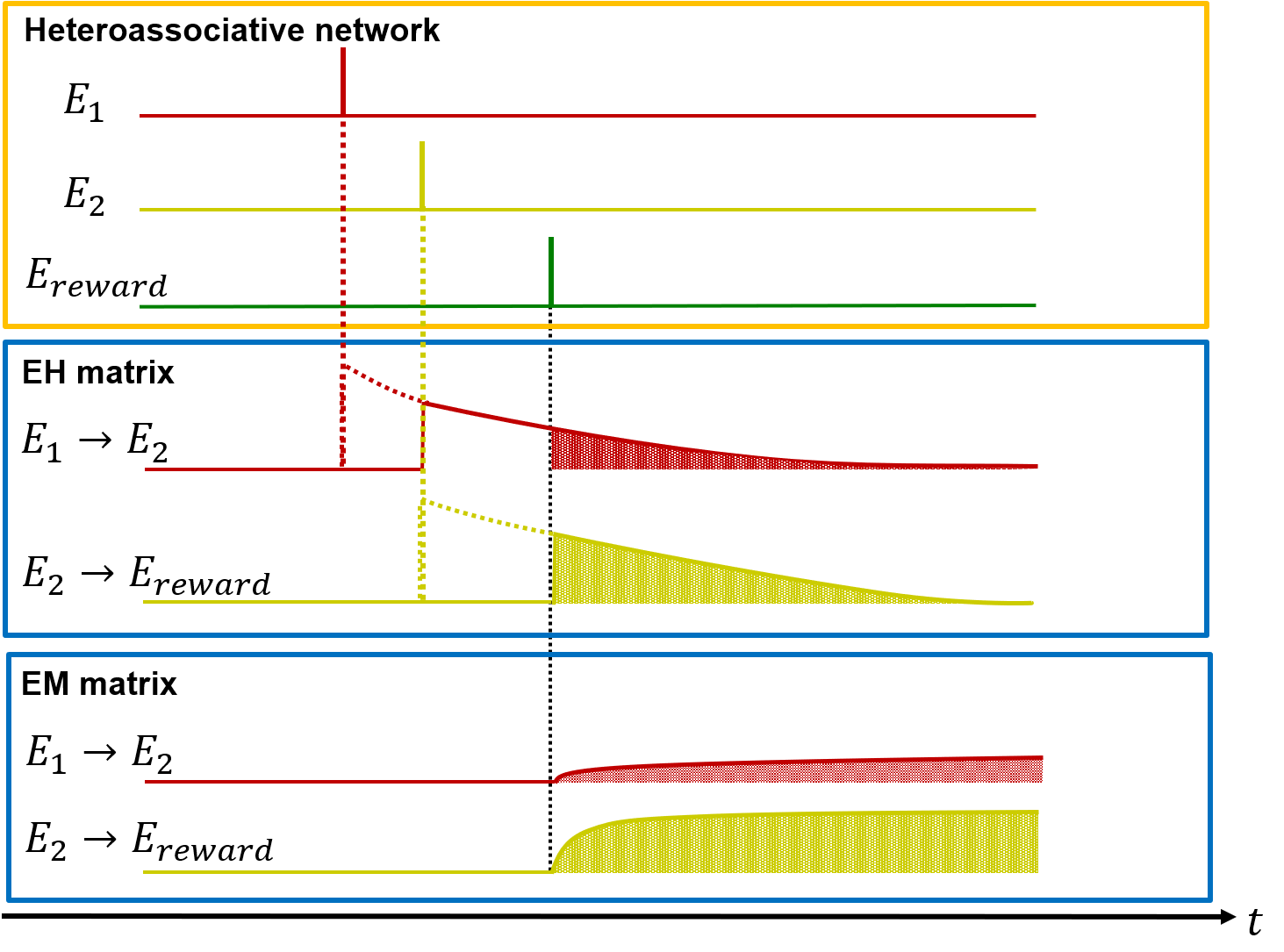}
  \end{center}
    \caption{Timing chart following the heteroassociative memory network with reinforced association between two coupled events by delayed reward $E_\text{reward}$. When $E_\text{reward}$ is delivered with a delay after temporally proximate events ($E_1$ and $E_2$), connections between $E_1$ and $E_2$ are potentiated.}
    \label{fig:subnet-timing-chart}
\end{figure}

The output of the heteroassociative memory network is temporarily held in the event-order holding (EH) matrix within a certain time after the event occurs. When a reward is given, associated event information in the EH matrix is fixed in the event-order memory (EM) matrix. Fig.~\ref{fig:subnet-timing-chart} depicts the timing chart that illustrates the reinforcement of firing patterns occurring on a millisecond timescale, even if the reward is delayed by a few seconds. This property is known as slow synaptic processes, which act as synaptic eligibility traces~\cite{klopf1982hedonistic, Sutton1998} or synaptic tags~\cite{Frey1997}.  A synaptic eligibility trace set a flag between pre- and postsynaptic neurons at a synapse that leads to a weight change only if an additional gating signal, such as a reward, is present. It is found in the hippocampus~\cite{Brzosko2015}, cortex~\cite{he2015}, and striatum~\cite{yagishita2014, shindou2019}. We introduce the EH matrix in which each matrix component works as eligibility trace. The cue-place cells fire at the timing of the corresponding events $E_1$ and $E_2$, and a temporal difference of these firing patterns is converted into event-order information by weighting eligibility variables between spikes in the EH matrix, as shown by the middle traces in Fig.~\ref{fig:subnet-timing-chart}. These eligibility variables exponentially decay over time. When a reward is fed, the eligibility variables are sampled, and they are converted into a variable representing the ordinal relationship between events in the EM matrix, as shown by bottom traces in Fig. ~\ref{fig:subnet-timing-chart}.

\subsection{Place cells}

Place cells fire when a mouse enters a particular place in an environment. In this study, the figure-eight maze includes $N_p$ place cells, and their receptive fields were predetermined. $N_p = x_{max} \times y_{max}$, where $x_{max}$ and $y_{max}$ are the maximum values in the $x$-axis and $y$-axis directions, respectively. A receptive field of a place cell is expressed by an activity bump, which is approximated by a 2-dimensional Gaussian distribution of firing rate. 
\begin{eqnarray}
    \label{eq:place-cell-gaus}
   &\nu^p_i(t) =
    \exp \left( - \frac{(x_i - x_c(t))^2 + (y_i - y_c(t))^2} {2 \sigma^2} \right)
\end{eqnarray}
where $(x_c, y_c)$ is the centroid of $i$-th place cell ($i = 1, 2, \cdots, N_p$), and $\sigma^2$ is the variance of the distribution.  The velocity $v$ and the head direction $\theta$ drives the location of the activity bump;
\begin{eqnarray}
    x_c(t + \delta t) = x_c(t) + v \cdot \delta t \cdot \cos \theta\\
    y_c(t + \delta t) = y_c(t) + v \cdot \delta t \cdot \sin \theta
\end{eqnarray}
where $\delta t$ is the unit time.

\subsection{Cue cells}

Cue cells fire in response to a non-spatial input derived by the signal of local view, which is called an environmental cue. Once an environmental cue signal is recognized, the firing rate of a $j$-th cue cell  ($j = 1, 2, \cdots, N_c$) is made, and this rate at a constant frequency.
\begin{eqnarray}
    &\nu^{c}_j(t) = \begin{cases}
        \mu_{th} & \text{if\ cue\ is\ recognized}\\
        0 & \text{otherwise}
    \end{cases}
\end{eqnarray}
where $N_c$ is the total number of an environmental cue. For simplification, one cue cell is associated with one environmental cue information. 

\subsection{Cue-place cells}

Cue-place cells integrate the neuronal activity of place and cue cells. The outputs of the place and cue cell corresponding to the $i$-th position and $j$-th cue stimulus, respectively, are combined in the cue-place cells. The firing rate of the cue-place cells $\nu_{k}^{cp}(t) \in \mathbb{R}^{N_e}$ is clipped at the certain value as follows:
\begin{eqnarray}
    \label{eq:conjunctive-cell}
    &\nu^{cp}_{k}(t) = \begin{cases}
        \nu^p_{i}(t) & \text{if}~\nu^c_{j}(t) \geq \mu_{th}\\
        0 & \text{otherwise}
    \end{cases}
\end{eqnarray}
where $N_e = N_p \times N_c$ is the total number of events, and $k = 1, 2, \cdots, N_e$. This means that the cue-place cell fires where the cue signal occurs.

\subsection{Heteroassociative network}

The heteroassociative network is a matrix $W(t)$ in which components $w_{pre,post}(t)$ are modulated between related cue-place cells. Firings of cue-place cells are seconds apart, even in related events. The STDP rule is only tied together limited to a postsynaptic firing that occurs within a few tens of milliseconds after a presynaptic firing~\cite{bi1998synaptic, mishra2016symmetric}. If a presynaptic firing is held several seconds, two temporarily distant firings can be linked. Neural activity is potentially held several tens of seconds by cue-holding neurons~\cite{hirabayashi2012} or time cells~\cite{macdonald2011hippocampal}. Here we assume that the neural activity of the cue-place cell is sustained by the cue-holding neuron-like mechanism. Thus, even if the postsynaptic cell fires a few seconds after the presynaptic cell fires, the synaptic connections are strengthened. However, we introduce an exponential decay function for synaptic potentiation. The synaptic weight modulation $F(\Delta t)$ between presynaptic and postsynaptic cells is as follows:
\begin{eqnarray}
    \label{eq:stdp-rule}
    F(\Delta t) = \begin{cases}
        \frac{A^{+}}{\tau_{+}} \exp (\Delta t/\tau_+) & (\Delta t > 0) \\
        0 & \left(\Delta t \leq 0 \right) 
    \end{cases},
\end{eqnarray}
where $\Delta t$ is the time difference between presynaptic and postsynaptic firing timings, and $A^+$ is the learning rate.

It would be mismanagement of resources if all neuron firing times were recorded to store the order in which events occur. So, we propose a method to extract event-order information. The synaptic weight matrix $W(t)$ of the heteroassociative memory network is modulated based on correlations between presynaptic and postsynaptic neurons. The temporal proximity between a featured event (post) and events that precede it (pre)  is retained as the magnitude of the synaptic weight $w_{pre, post}$ ($pre = 1, 2, \cdots, N_e, post = 1, 2, \cdots, N_e$). The component of $W(t)$ is described below:
\begin{eqnarray}
    W(t) = \left(
    \begin{array}{cccc}
      w_{1, 1}(t) & w_{1, 2}(t) & \ldots & w_{1, N_e}(t) \\
      w_{2, 1}(t) & w_{2, 2}(t) & \ldots & w_{2, N_e}(t) \\
      \vdots & \vdots & \ddots & \vdots \\
      w_{N_e, 1}(t) & w_{N_e, 2}(t) & \ldots & w_{N_e, N_e}(t)
    \end{array}
    \right)
\end{eqnarray}
\begin{eqnarray}
  &w_{pre,post}(t) = \displaystyle \int^{t}_{t^{\prime}} \nu^{cp}_{pre}(t^{\prime}) \nu^{cp}_{post}(s) F(s - t^{\prime}) ds\\
  &t^{\prime} = t - T_s
  \label{eq:ham-rule}
\end{eqnarray}
where $T_s$ is the period for seeking event pairs. $\nu^{cp}_{post}$ is the neural activity of the post event, and $\nu^{cp}_{pre}$ is the neural activity of the pre event.

\subsection{Event-order holding matrix}

All the synaptic weights expressed by the matrix $W(t)$ of the heteroassociative memory network are fixed in the EM matrix via the EH matrix. If all events throughout the learning period are stored, a huge amount of computational resources are consumed to express the order of the events. To solve this problem, we proposed the EH matrix using a synaptic eligibility trace~\cite{izhikevich2007solving}. 

The sequence information for each event, which decays over time, represented by a heteroassociative memory network, is summed up in the EH matrix. The matrix $C(t) = (c_{pre, post}(t)) \in \mathbb{R}^{N_e \times N_e}$ of the eligibility variable is stimulated by $W$.
\begin{equation}
    \frac{dc_{pre, post}(t)}{dt} = - \frac{c_{pre, post}(t)}{\tau_c } + w_{pre, post}(t)
\end{equation}
The eligibility variable $c_{pre, post}(t)$ exponentially decays with the time constant $\tau_{c}$. The initial value $C(0)$ is an $N_e \times N_e$ zero matrix.

\subsection{Event-order memory matrix}

The order information of past events is transferred from the EH matrix to the EM matrix $M = (m_{pre,post}) \in \mathbb{R}^{N_e \times N_e}$. We introduce a mechanism to store events synchronized with the reward timing in the memory. The component $m_{pre, post}$ represents the strength of the relationship between the events $E_{pre}$ and $E_{post}$. 
\begin{equation}
M(t + \delta t) \leftarrow M(t) + d(t) C(t) 
\end{equation}
When a reward is given at $t_{reward}$, the eligibility variable $C$ allows a change of $M(t)$.  The initial value $M(0)$ is an $N_e \times N_e$ zero matrix. This change is gated by the reward-associated value $d$ of dopamine. 
\begin{equation}
\frac{dd(t)}{dt} = - \frac{d(t)}{\tau_d} + \delta(t - t_{reward})\\
\end{equation}
where $d(t)$ exponentially reduces with time constant $\tau_d$, and $\delta(t)$ is the delta function. If $t = t_{reward}$, $\delta = 1$; otherwise $\delta = 0$.

\subsection{Recall scheme using synaptic weights stored in the EM matrix}

Herein, we describe how to recall event-order information from the EM matrix $M(t)$. Let us consider recall procedures to extract events associated with impetus inputs event according to the order of stored events. A recall matrix $A^k$ is prepared to recall the event-sequence  information stored in the EM matrix $M(t)$, where $k$ is an iteration number of recall procedures. The recall matrix $A^{k}$ is formed by operating $M$ from the right on the initial matrix $A^0$. By repeatedly applying the EM matrix $M(t)$ to the recall matrix, the events are recalled in the order starting from the impetus input $I_{s}$. Recall procedures are shown in the following equations:
\begin{eqnarray}
    &A^k \leftarrow A^{k-1} \cdot M(t)\\
    \label{eq:recall-noise}
    &A^k = \left(
    \begin{array}{cccc}
      a^k_{1, 1} & a^k_{1, 2} & \ldots & a^k_{1, N_e} \\
      a^k_{2, 1} & a^k_{2, 2} & \ldots & a^k_{2, N_e} \\
      \vdots & \vdots & \ddots & \vdots \\
      a^k_{N_e, 1} & a^k_{N_e, 2} & \ldots & a^k_{N_e, N_e}
    \end{array}
    \right)\\
    &	a^0_{q,r} = \begin{cases}
        1 & \text{if}~q = r = s\\
        0 & {\text{otherwise}}
    \end{cases}\\
    &\{\forall k, q, r, \text{if}~a^k_{q, r} \leq \mu_{th}\  \text{then}\  a^k_{q, r} = 0\}
    \label{eq:recall-repeat}
\end{eqnarray}
where $s (= 1, 2, \cdots, N_e$) is a component of the impetus input to recall $I_{s}$. If the input value exceeds the threshold value $\mu_{th}$, $a^k_{q, r}$ equals the input value; otherwise, $a^k_{q, r}$ equals $0$. Repeating this recall process starting from the impetus input $I_s$, temporally distant events can be recalled.

\section{Experimental results}
\subsection{Experimental setup}

\begin{figure*}[h]
  \begin{center}
    a. Route recall task\\
    \includegraphics[width=0.7\linewidth, keepaspectratio, clip]{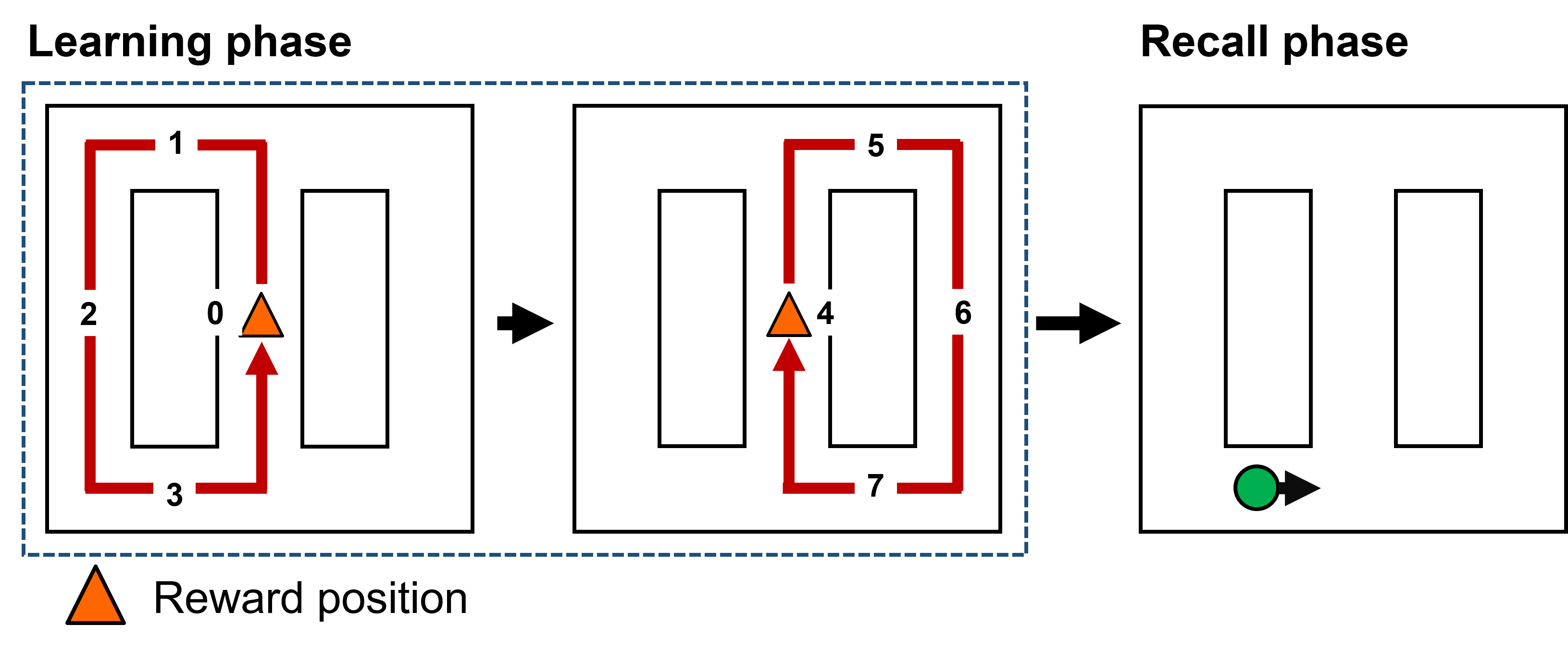}\\
    b. Reward recognition task\\
    \includegraphics[width=0.7\linewidth, keepaspectratio, clip]{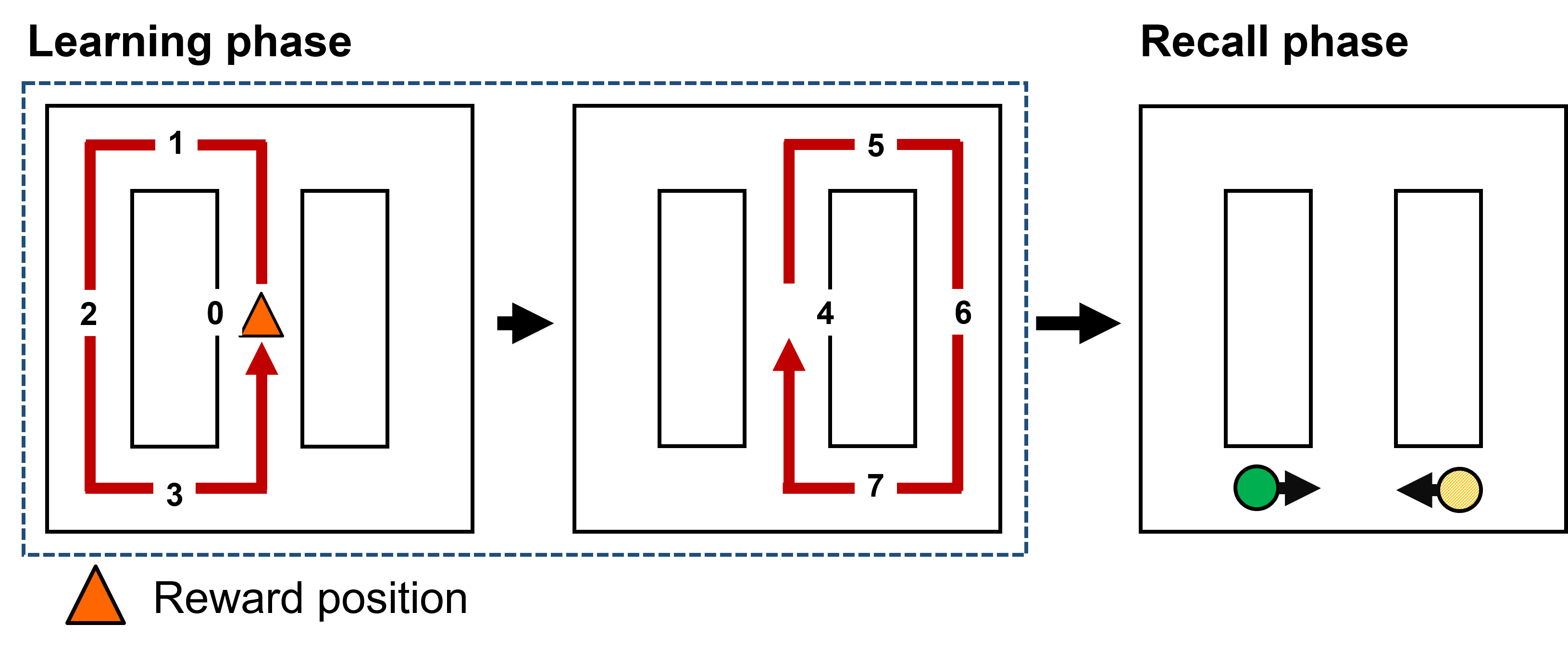}\\
    c. Constructed-route retrieval task\\
    \includegraphics[width=0.7\linewidth, keepaspectratio, clip]{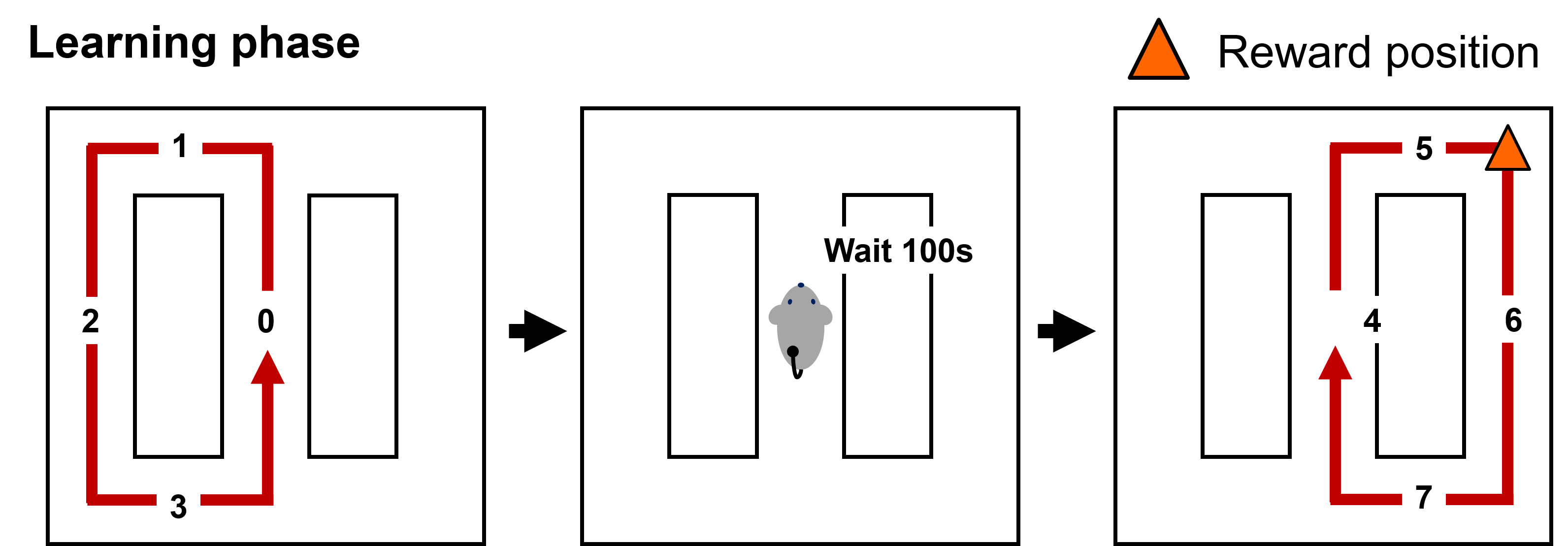}
  \end{center}
    \caption{Route overview. (a)  Route recall task. In the learning phase, the mouse moves the counterclockwise along the aisle number, returns to the initial point, and then moves clockwise. The reward is fed when the mouse returns to the initial location. In the recall phase, an impetus input $I_s$  (green circle) is entered and the related locations will be called sequentially. (b) Reward recognition task. In the learning phase, the mouse moves the counterclockwise along the aisle number, returns to the initial point, and then moves clockwise. The reward is fed only after moving counterclockwise or clockwise. (c) Constructed-route retrieval task. The virtual mouse stays a hundred of seconds at the central aisle after the counterclockwise movement and then moves the clockwise. The reward is fed at the right corner in the clockwise movement.}
    \label{fig:experiment-route-overview}
\end{figure*}

We conducted computer simulations in which a virtual mouse traveled in the figure-eight maze, as shown in Fig.~\ref{fig:experiment-route-overview}. Each aisle is numbered in the order in which it moves. The mouse moves counterclockwise from the center aisle, returns to its initial position, and then moves clockwise. Since the center aisle was visited twice, two numbers were assigned.

The map size $x_{max} \times y_{max}$ to be simulated was $50 \times 50~\rm{cm}$. In the maze, 2500 place cells were assigned. $\mu_{th}$ = 10. A moving speed $v$ of 8 -- 10~\rm{cm/s} was set. Movable directions were RIGHT, UP, LEFT and DOWN. The head direction was considered as an environmental cue since any object seen when the head is in an orientation is a cue. Also, the reward was used as an environmental cue for the same reason. The number of cues $N_c$ is 5. The number of events $N_{e}$ prepared was 12,500. The unit time $\delta t$ was set at $0.1$~s.  The time constant $\tau_+$ is set at 4~s. Although it takes time to move a virtual mouse, if the time window of the learning function $F(\Delta t)$ is long enough, the activities of the cue-place cells are connected along the route of the excursion. We note that $T_s$ is 2~s throughout the simulation. The time constant $\tau_{c}$ for the EH matrix is 8~s and the time constant $\tau_d$ for the EM matrix is  3~s. Table~\ref{tbl:block-num} shows the number of neurons and synapses in the network in our experiments.
\begin{table}[ht]
    \begin{center}
        \caption{Numbers of neurons or size of matrices in our experiments} 
        \label{tbl:block-num}
        \begin{tabular}{|p{0.45\linewidth}|p{0.45\linewidth}|}
            \hline
            Block name & Number \\ \hline \hline
            Speed cell & 1 \\
            Head direction cell & 1 \\
            Place cell & 2,500 ($50 \times 50$) \\
            Cue cell & 5 \\
            Cue-place cell & 12,500 ($2,500 \times 5$) \\
            Heteroassociative memory network & 250,000 ($12,500 \times 20^*$) \\
            Event-order holding matrix & 156,250,000 ($12,500 \times 12,500$) \\
            Event-order memory matrix & 156,250,000 ($12,500 \times 12,500$) \\
            \hline
        \end{tabular}
    \end{center}
      * $T_s / \delta t$ = 20
\end{table}

In this study, three tasks were conducted. The first experiment was a path recall task. We verified whether it was possible to recall the moved routes in the order in which they were experienced using the proposed model. The second experiment was a reward recognition task. We verified whether the rewarded route could be recognized if the reward was given differently depending on the route traveled. The third task was a constructed-route retrieval task. We tested whether the mouse could integrate the traveled routes to reach the reward when it traveled the first route, took a break, and then traveled the next route.

\subsubsection{Route recall task}

During the learning phase, the virtual mouse proceeds in the aisle number order $0 \rightarrow 1 \rightarrow 2 \rightarrow 3 \rightarrow 0$, and then $4 \rightarrow 5 \rightarrow 6 \rightarrow 7 \rightarrow 4$. The initial location was the center aisle. The reward was given when the mouse returns to the initial location. In the recall phase, which is after the learning phase, an impetus input $I_s$ to recall was set, and the recall procedure was executed. $I_s$ is defined by $x_{sx}$, $y_{sy}$ and $c_{sc}$, where $sx \in [1, x_{max}], sy \in [1, y_{max}]$, and $sc = 1, 2, \cdots, N_c$.

\subsubsection{Reward recognition task}
\label{sec:reward-recognition-task-setup}

In the route recall task, the virtual mouse traveled counterclockwise and clockwise along the aisle numbers during the learning phase. The initial location was $(25, 25)$. We prepared two different timings for giving rewards. In the first timing, the virtual mouse was rewarded in the center aisle after the counterclockwise excursion. In the second timing, the reward was given after the clockwise excursion. 

\subsubsection{Constructed-route retrieval after two single-loop tasks}
\label{sec:constructed-route-retrieval-task-setup}

Fig.~\ref{fig:experiment-route-overview}c depicts an overview of the route used in the constructed-route retrieval task. The virtual mouse took 100~s break after completing the counterclockwise route, and then it traveled the right aisle clockwise. No reward is fed in the counterclockwise movement, but it is fed only when the point $(45, 45)$ in the clockwise movement is passed.

In the recall phase, the impetus input was given to the lower-left point ($I_{2}(20, 10, \text{RIGHT})$) in the counterclockwise route.

Numerical simulations were performed by a multi-core CPU (Intel Core i7 6700K) with a graphic processing unit (NVIDIA, GTX 1050Ti ).

\subsection{Route recall task}

\begin{figure}[h]
  \begin{center}
    \includegraphics[width=\linewidth, keepaspectratio, clip]{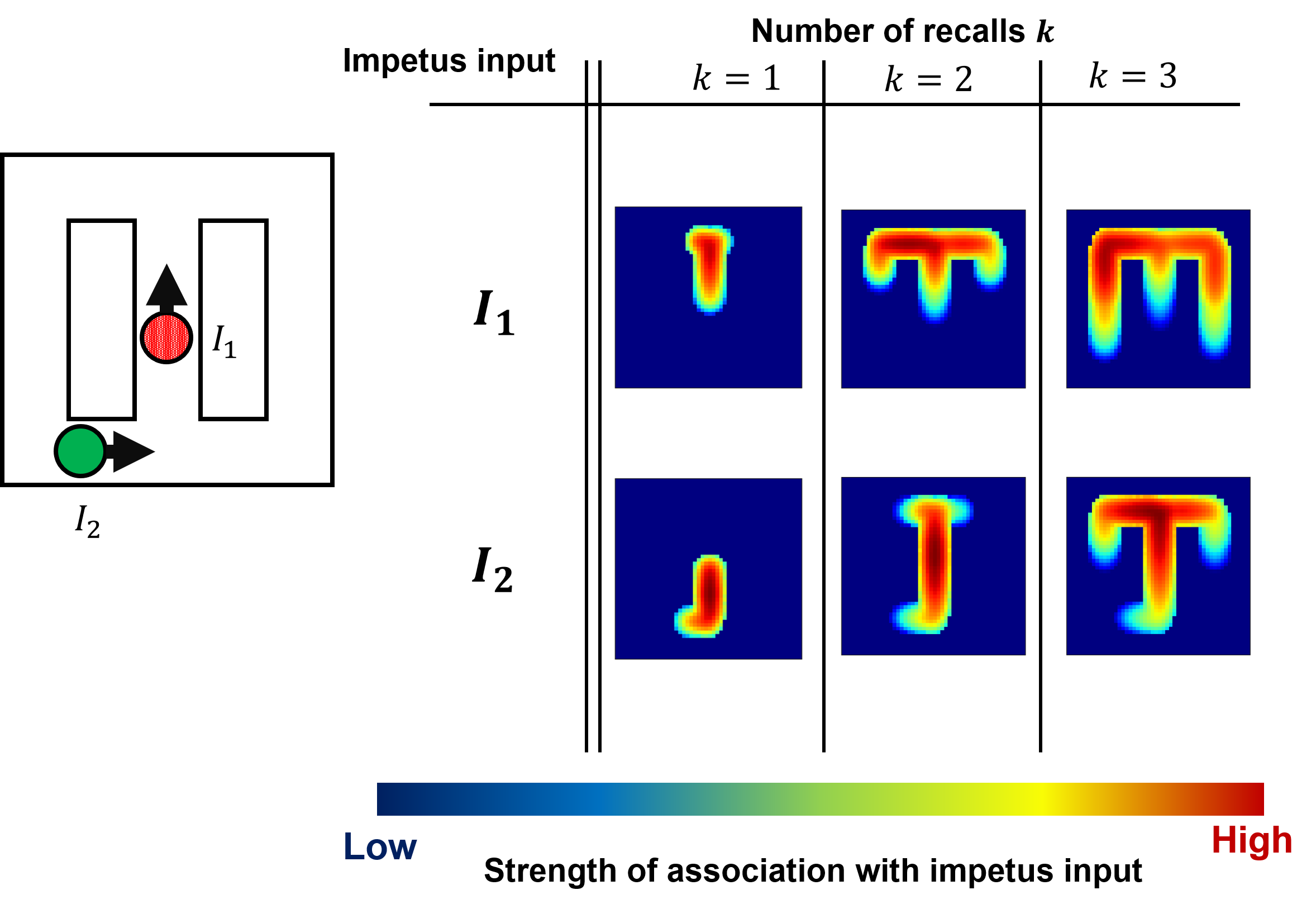}
    \caption{Route recall task results}
    \label{fig:route-recall-task-result}
  \end{center}
\end{figure}

Fig.~\ref{fig:route-recall-task-result} depicts the value of $A^{k}$ displayed as a heat map. The matrix $A^k$ is obtained by repeatedly applying the recall procedures starting from the impetus input $I_s$, which is $I_{1}(25, 25, \text{UP})$ or $I_{2}(20, 10, \text{RIGHT})$. The heat map represents the recalled spatial receptive field. The recalled procedure was iterated up to three times at each event. The number of events acquired by recalls increased with an increase in the number of recalls. Starting from the impetus input $I_{s}$, the routes taken by the $I_{s}$ associated events were recalled. The length of the recalled routes increases with an increase in the number of recalls $k$. The target event location was recalled by tracing the events in the order of temporal proximity to the trigger event.

When the impetus input was $I_{1}$, as shown in Fig.~\ref{fig:route-recall-task-result}, the counterclockwise and clockwise routes were equally recalled at the 3-way junction in the upper center. Since the sequence information of the learned routes was degraded in the EH matrix with time, the relationship strength obtained by recall in the clockwise route should be higher than that in the counterclockwise route. However, the same relationship strength was obtained for both routes. The equality in relationship strength is because the reward was given when the head direction changed in the route recall task, and the learning result was fixed in the EM matrix before the time decay of the learned route became noticeable.

\subsection{Reward recognition task}

\begin{figure}[h]
  \begin{center}
    a\\
    \includegraphics[width=\linewidth, keepaspectratio, clip]{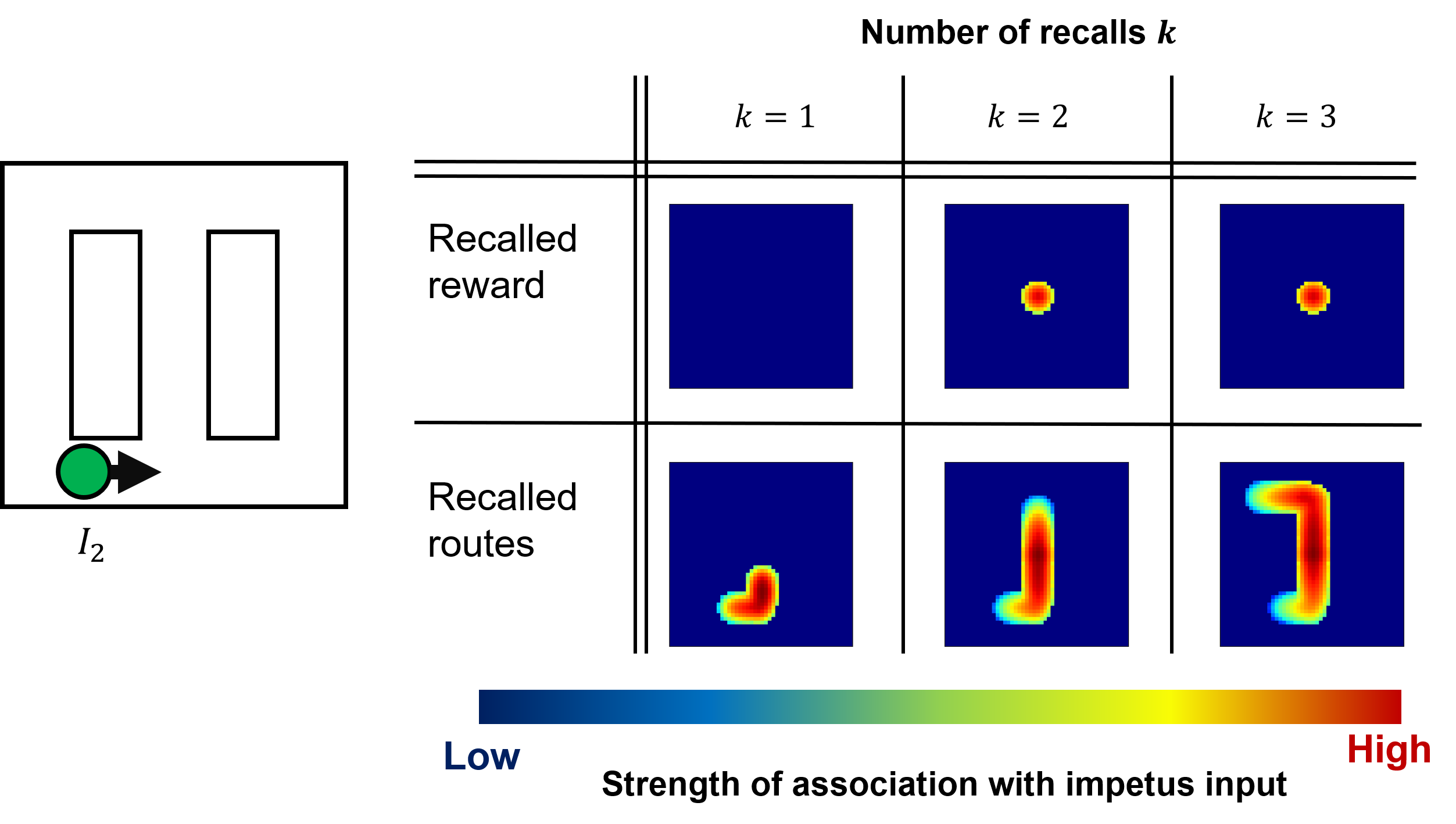}
    b\\
    \includegraphics[width=\linewidth, keepaspectratio, clip]{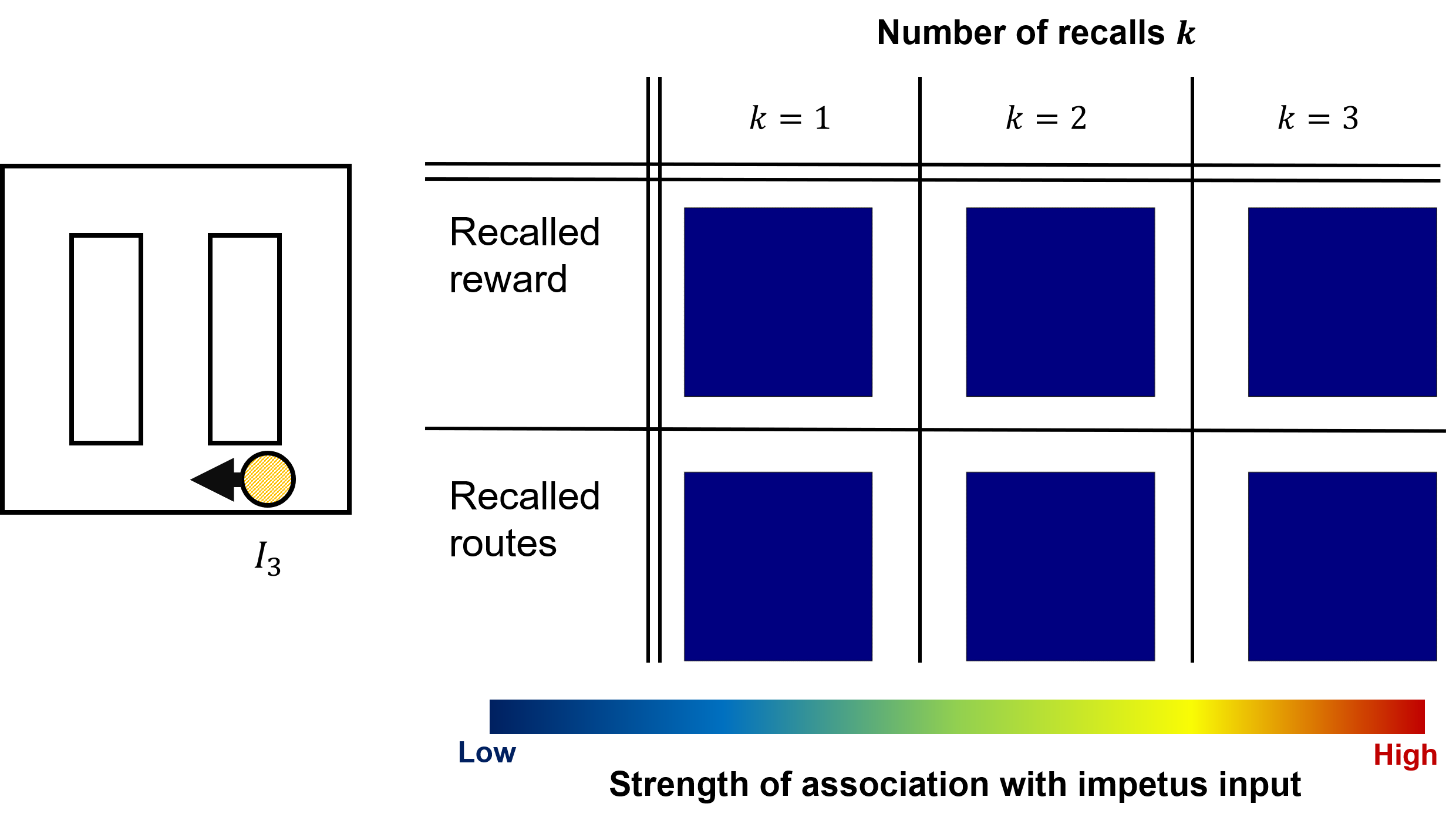}
    \caption{Experimental results of reward recognition task (1). (a)The impetus input is $I_1$, which is an event happened in the counterclockwise route. (b) The trigger event is $E_2$, which is an event on the clockwise route.}
    \label{fig:reward-recognition-task-result-left-rewarded}
  \end{center}
\end{figure}

In the reward recognition task, a reward was given only during either the counterclockwise or clockwise excursion. For recalling the reward location, the recall steps were repeated after the excursions. The following two inputs: $I_2$ and $I_3$, were prepared as trigger events, as shown by the green and yellow circles in Figs.~\ref{fig:reward-recognition-task-result-left-rewarded}a and \ref{fig:reward-recognition-task-result-left-rewarded}b, respectively. Further, $I_{2}(20, 10, \text{RIGHT})$ occurred in a counterclockwise route, whereas $I_{3}(30, 10, \text{RIGHT})$ occurred in a clockwise route. Although the reward was not located near $I_{2}$, by repeating the recall procedures from the trigger event, the reward location ans route to the reward appeared. Thus, when $I_{2}$ is given, the reward and route to the reward were successfully retrieved after a couple of recall steps $k = 2$ and $3$, as shown in Fig.~\ref{fig:reward-recognition-task-result-left-rewarded}a. However, when $k = 1$, the reward was not recalled. 

When $I_{3}$ is given, both the reward and the route to the reward were not recalled, as shown in Fig.~\ref{fig:reward-recognition-task-result-left-rewarded}b. So neither the reward nor route to the reward was recalled at any number of recalls $k$. This is because no reward is given in the clockwise excursion. Therefore, in the EM matrix, the reward is neither directly nor indirectly associated with $I_{3}$ on the clockwise path. The virtual mouse reaches the reward location in the central stem during mind travel only if an impetus input to remember is on the counterclockwise excursion in which the reward was gained.

\begin{figure}[h]
  \begin{center}
    a\\
    \includegraphics[width=\linewidth, keepaspectratio, clip]{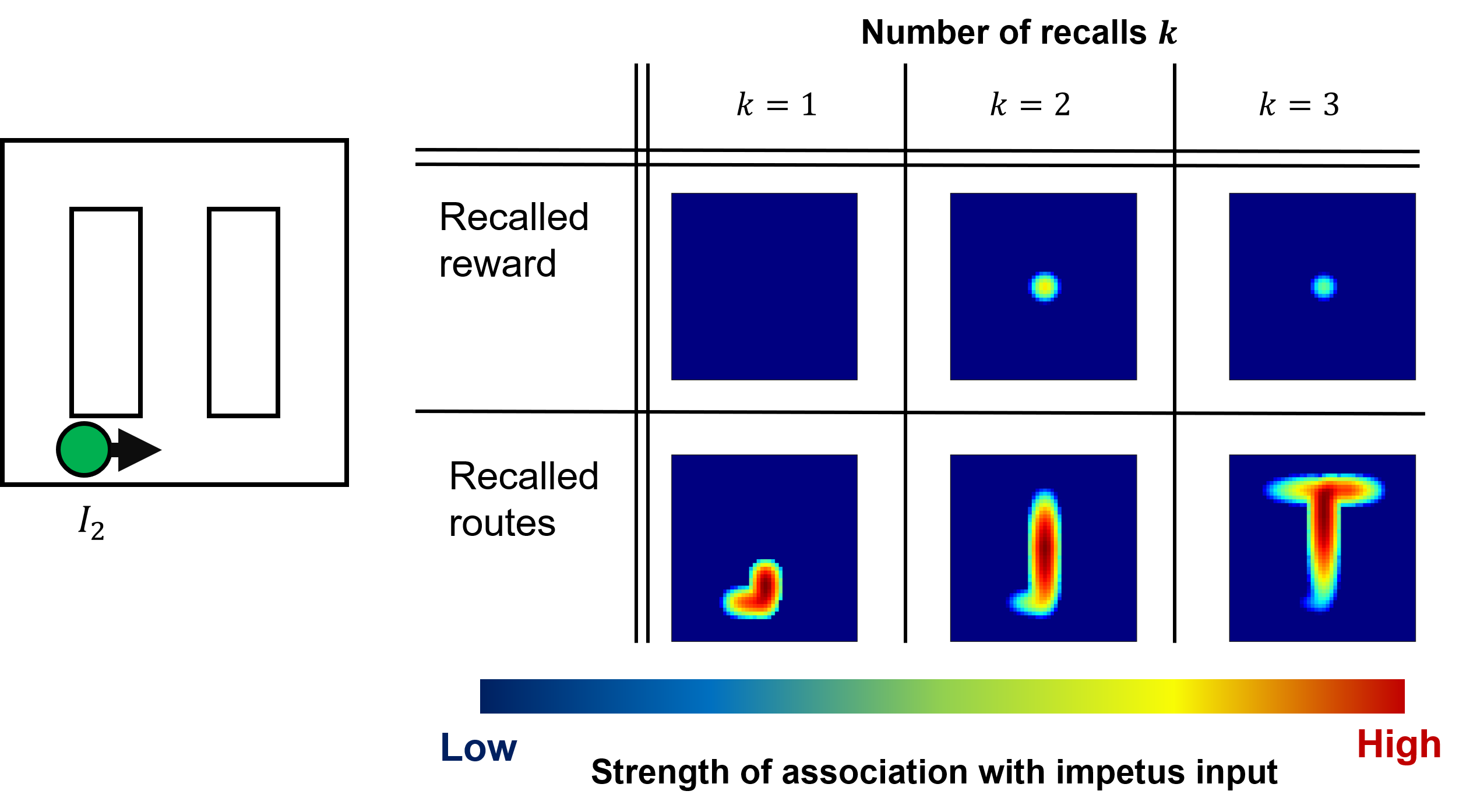}
    b\\
    \includegraphics[width=\linewidth, keepaspectratio, clip]{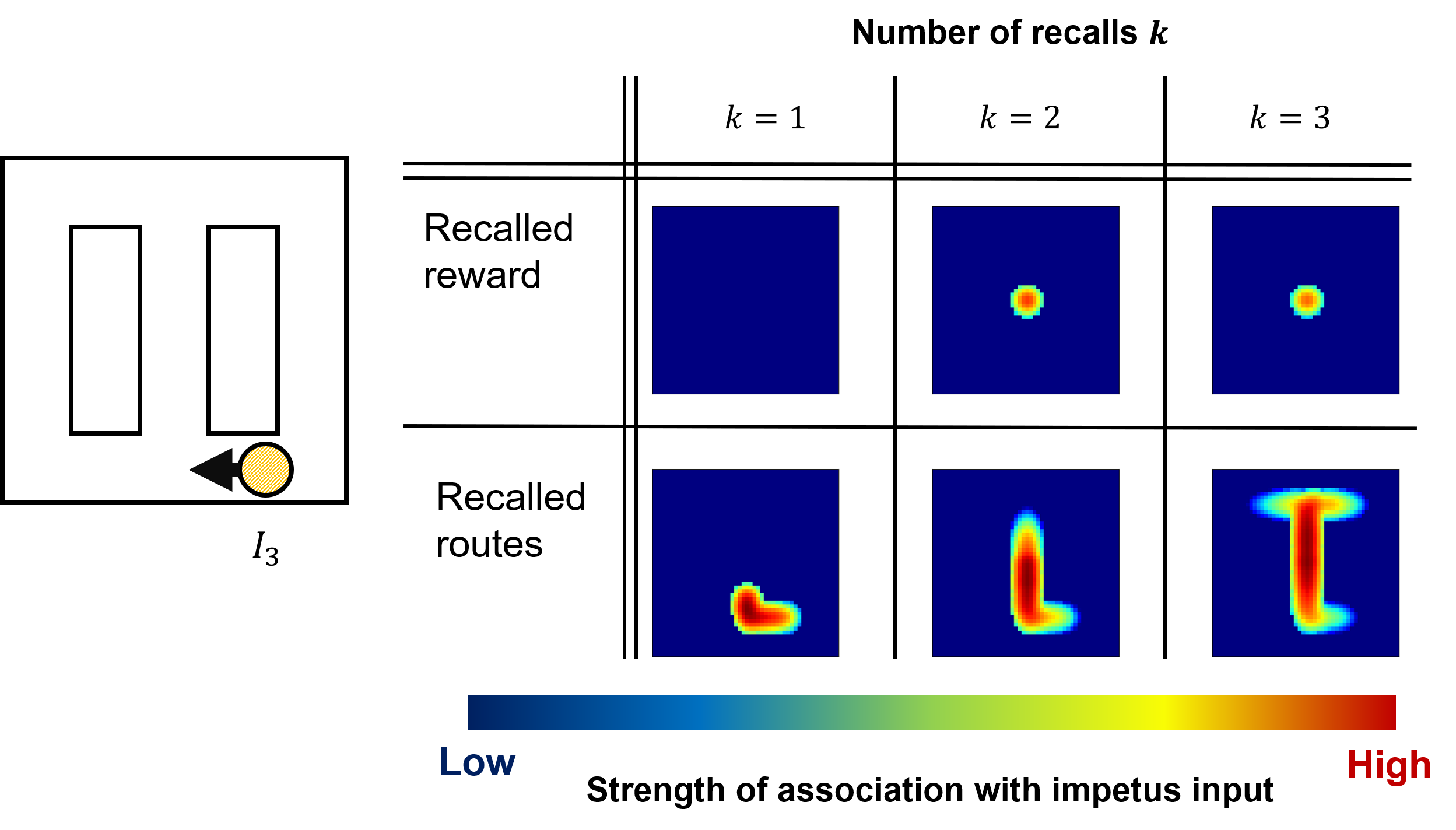}
  \end{center}
   \caption{Experimental results of reward recognition task (2). (a) The impetus input is $I_2$, which is an event on the counterclockwise route. (c) The trigger event is $E_3$, which is an event on the clockwise route.}
    \label{fig:reward-recognition-task-result-right-rewarded}
\end{figure}

Fig.~\ref{fig:reward-recognition-task-result-right-rewarded} depicts the heat map in the second reward recognition task in which the reward is fed only after moving clockwise. Figs.~\ref{fig:reward-recognition-task-result-right-rewarded}a and \ref{fig:reward-recognition-task-result-right-rewarded}b show the experimental results triggered by $I_{2}$ and $I_{3}$, respectively. In both figures, the reward was recalled when the number of recalls $k = 2, 3$. However, there is a color difference in the heat map. This color difference is thought to be caused by the time difference between learning the route and being rewarded. More specifically, it is because the counterclockwise route information was time-attenuated by the EH matrix, and it is accumulated in the EM matrix more than the clockwise route information. Additionally, in the 3-way junction route in the upper center, which was recalled when the number of recalls $k = 3$, the heat map of the route to the left is smaller than that to the right. This difference in the heat map size is because the information on the movement order of the counterclockwise route is accumulated in the EM matrix after time attenuation. Accordingly, it is thought that the clockwise route has a stronger relevance to the reward than the counterclockwise route.

\subsection{Constructed-route retrieval task}

\begin{figure}[h]
  \begin{center}
    \includegraphics[width=\linewidth, keepaspectratio, clip]{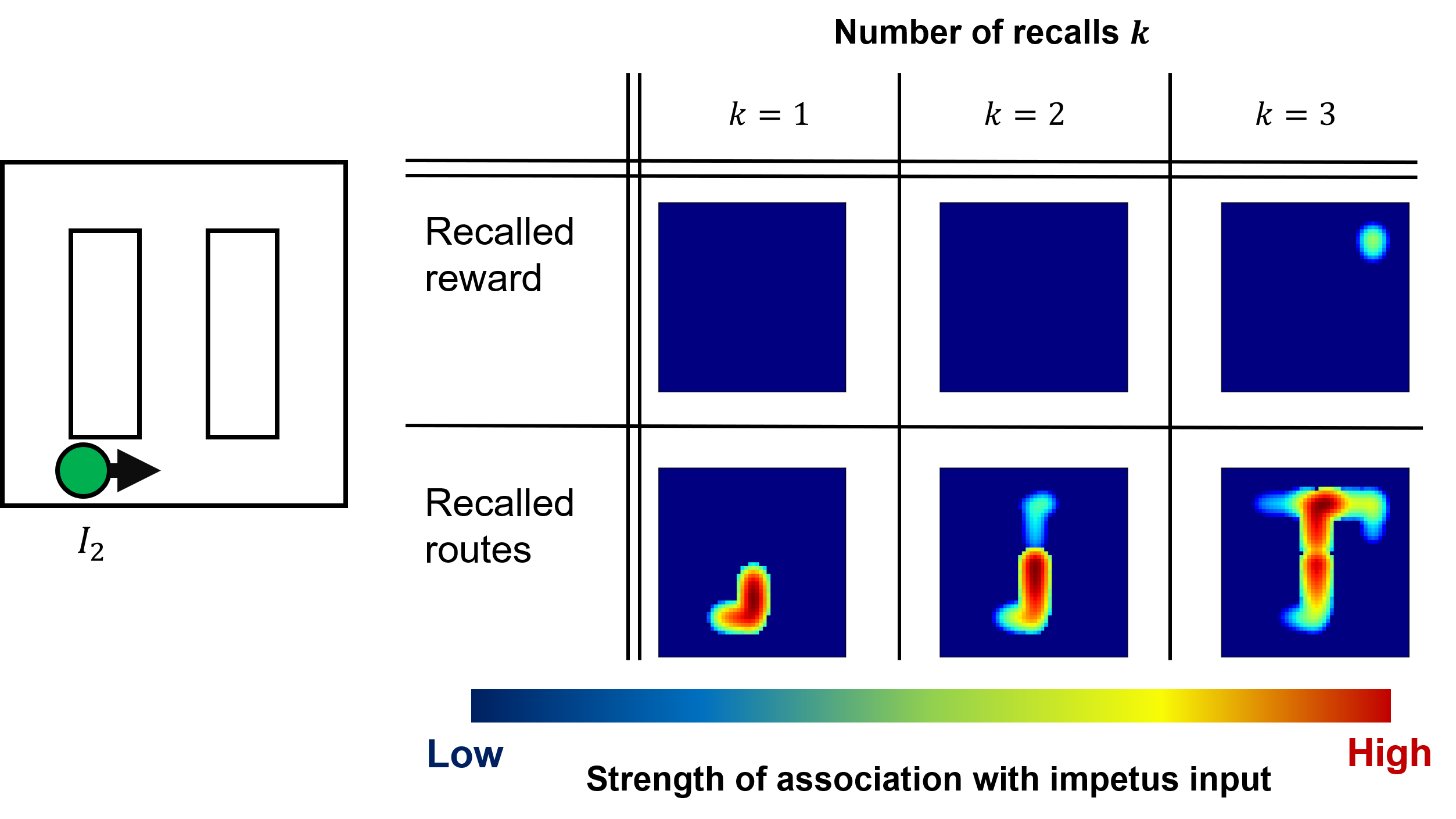}
  \end{center}
   \caption{Experimental results of the constructed-route retrieval task. The impetus input is $I_2$, which is an event on the counterclockwise route.}
    \label{fig:constructed-route-retrieval-task-result}
\end{figure}

Fig.~\ref{fig:constructed-route-retrieval-task-result} depicts the heat map in the constructed-route retrieval task. This heat map was obtained by repeatedly applying the recall procedures starting from the impetus input $I_{2}(20, 10, \text{RIGHT})$. The clockwise route was learned with a break after the counterclockwise excursion was conducted. For $k = 2$, the central aisle has a bright color. Due to the waiting time of 100~s, the sequence information of the events related to the counterclockwise route expressed on the EH matrix was attenuated with time and then accumulated in the EM matrix. This process is considered to have caused this color difference in the heat map. This result indicates that the experience can be integrated to indirectly associate the trigger event with the reward, even if there is a time lag before obtaining the reward.

A virtual mouse could reach the point where the reward was given in the clockwise route when $k = 3$, even when recalling from the impetus input $I_{2}$ in the counterclockwise route. For $k = 3$, the heat map of the right route at the T-junction on the central stem is brighter than that of the left route. The difference in brightness is because the reward was given at the right corner on the clockwise route, and the event order during the clockwise route was less attenuated. It is considered that this phenomenon occurred because it was judged that the behavior that was close in time to the timing of receiving the reward in this model was more valuable.

We can expect to integrate the routes that the virtual mouse had traveled so far and associate it with the position where the reward was given, even if the virtual mouse recalls from positions on counterclockwise route that was not rewarded.

\section{Discussion}

For temporally close events, the Hebb’s and STDP rules allow to potentiate synaptic connections between neural assemblies of the events. However, if there is a relationship between events A and B, but with a gap in time, it will be difficult to link that relationship by directly enhancing the synaptic connections between the neuronal ensembles representing the events A and B. Also, it will be difficult to augment the causal relationship between place cells that determine action and neural activity representing the action. This difficulty is because if a reward is obtained as a result of the action, the neural activity that represents the action is likely to have ended before the reward is obtained. Therefore, those temporally distant neural ensembles can be linked together because of learning. To address this problem, we implemented two stages of learning: one for temporally distant events and the other for reward-related learning.

In the route recall task, the recall mechanism was devised based on the idea that the target event is indirectly reached by tracing the events that are directly related to the recall trigger event in order. The time series information of the events is converted into the event-order information, and it is held in the event-order memory matrix. Since the relationship between directly related events is maintained in the EM matrix, it is reasonable to recall both routes when the rat passes through both the counterclockwise and clockwise routes.

In the constructed-route retrieval task, it was still possible to find a route from the counterclockwise route to the rewarded route, even when the counterclockwise route was followed by the clockwise route with a 100 s waiting period in between. By incorporating an event holding function, it is possible to link trigger events and reward locations that are separated in time.

We employed a heteroassociative network that encodes the sequential memory of events into a matrix. This memory of events is then transferred to the matrix of long-term memory. The hippocampal CA3 is considered an autoassociative memory because of its recurrent connections. However, the hippocampal CA3 has been also considered a heteroassociative memory because of the phase precession of place cells~\cite{Jensen1996} \cite{Lisman1999}. Several network models have been proposed for heteroassociative memory~\cite{Sompolinsky1986, Levy1996}. de Camargo et al.~\cite{deCamargo1998} have proposed the hippocampal CA3 network model for pattern sequence recall based on the heteroassociative network. In our model, the pattern sequences are given according to the joint information about the place and associated cue when exploring the maze. Thus, the events are recalled in the order in which they occurred, and the path from a certain point in the maze to the reward can be stored. 

In the heteroassociative network, the relatively long time constant of synaptic potentiation is assumed. The physiological time constant of the STDP function is several tens of milliseconds. In the medial temporal lobe, persistent firing neurons, cue-holding cells~\cite{hirabayashi2012, naya2003} and time cells~\cite{macdonald2011hippocampal} have been found. In a pair association memory task, neural activity of a cue-holding cell induced by a cue stimulus sustains during a couple of seconds of the delay period~\cite{hirabayashi2012, naya2003}. Time cells fires after an elapsed period from a cue input~\cite{macdonald2011hippocampal}. Moreover, all of these cells continue to fire for several tens of seconds after stimulation. Such sustaining firings potentially contribute to linking temporally distant events and informing the timing of a series of events. There is a physiological contradiction in synchronization between two distant events, but it can be resolved by introducing time or cue-holding cells.

\section{Conclusion}
We proposed an entorhinal cortex and hippocampus neural network model with a long-term memory matrix to encode event-order information and to consolidate that event sequences in the matrix. The route to the reward is recalled by successive repetition of the recall steps in the mind travel. Such recall mechanisms for goal-oriented behavior are beneficial for mobile service robots that perform autonomous navigation. Further, a neural network that remembers the sequential relationships of related events in a single trial leads to a system that easily learns from experience.

\section*{Acknowledgment}
This research is supported by JSPS KAKENHI Grant Numbers 20K21819. This research is also based on results obtained from a project, JPNP16007, commissioned by the New Energy and Industrial Technology Development Organization (NEDO).

\bibliographystyle{IEEEtran}
\bibliography{ieee_access}

\end{document}